\documentclass[12pt]{article}
\usepackage{amssymb,amsmath}
\usepackage{epsfig}
\def\N{{\cal N}}
\def\L{{\cal L}}
\def\E{{\cal E}}
\def\S{{\cal S}}
\def\O{{\cal O}}
\def\D{{\cal D}}
\newcommand{\axs}{AdS_5\times S^5}
\def\r{\rho}
\def\a{\alpha}
\def\b{\beta}
\def\d{\delta}
\def\g{\gamma}
\def\G{\Gamma}
\def\s{\sigma}
\def\t{\tau}
\def\x{\xi}
\def\m{\mu}
\def\n{\nu}
\def\k{\kappa}
\def\p{\partial}
\def\f{\phi}
\def\o{\omega}
\def\e{\eta}
\def\ep{\varepsilon}
\def\l{\lambda}
\def\th{\theta}
\def\rb{\right}
\def\lb{\left}

\newcommand{\eq}[1]{\begin{equation} #1 \end{equation}}
\newcommand{\al}[1]{\begin{align} #1 \end{align}}
\newcommand{\ml}[1]{\begin{multline} #1 \end{multline}}

\begin{document}
\begin{center}
{\bf{\Large  Rotating Strings with B-field} \\
\vspace*{.35cm}
}

\vspace*{1cm}

R.C. Rashkov\footnote{e-mails: rrachkov@sfu.ca; rash@phys.uni-sofia.bg,
on leave of absence from Dept. of Physics, Sofia University, 1164 Sofia,
Bulgaria}
and K.S. Viswanathan\footnote{e-mail: kviswana@sfu.ca}\\

\ \\
Department of Physics, Simon Fraser University \\
Burnaby, BC, V5A 1S6, Canada
\end{center}

\vspace*{.8cm}

\begin{abstract}
Some of the recent important developments in understanding
string/ gauge dualities are based on the idea of highly symmetric
motion of ``string solitons'' in $AdS_5\times S^5$ geometry originally 
suggested by Gubser, Klebanov and Polyakov. In this paper we study symmetric
motion of short strings in the presence of antisymetric closed string
B field. We compare the values of the energy and the spin in the
case of non-vanishing B field with those obtained in the case of B=0. 
The presence of NS-NS antisymmetric field couples the fluctuation
modes that indicates changes in the quantum corrections to the energy spectrum.
\end{abstract}

\vspace*{.8cm}

\section{Introduction}

The main developments and research efforts in string theory in the last
years were focused on the understanding of string/gauge duality and
especially AdS/CFT correspondence. AdS/CFT correspondence is based on
the conjecture that type {\bf II B} string theory in $AdS_5\times S^5$
background with large number of fluxes turned on $S^5$ is dual to
four dimentional $\N=4$ supersymmetric Yang-Mills theory (SYM).
 Actually this is the best studied case where the spectrum of a
state of string theory on $\axs$ corresponds to the spectrum of single
trace operators  in the gauge theory. The great interest in
AdS/CFT correspondence is inspired by the simple reason that it can be
used to make predictions about $\N=4$ SYM at strong coupling.

Until recently this conjecture has been mostly tested in the
supergravity approximation. In this case it was found that the
supergravity modes on $\axs$ are in one to one correspondence with the
chiral operators in the gauge theory. However, this restriction means
that the curvature is small and all $\alpha' R$ corrections are
neglected ($R$ is the radius of $AdS$). These  trancated
considerations are not enough to verify the correspondence in its full
extent and to extract useful information about the gauge theory at
strong coupling. It is important therefore to go
beyond the supergravity limit and to consider at least  $\alpha'
R$ correction.

The recent progress in string propagation in pp-wave background
attracted much interest \cite{b1, b2, b3}. This background possesses
at least two very attractive features: it is maximally supersymmetric
and at the tree level Green-Schwarz superstring is exactly solvable
\cite{mets1, metsT1}. Since this background can be obtained by taking the
Penrose limit \cite{penr} of $\axs$ geometry, it is natural then to
ask about AdS/CFT correspondence in these geometries. In an important
recent development, Berenstein, Maldacena and Nastase (BMN) \cite{bmn}
have been able to connect the tree level string theory and the dual SYM
theory in a beautiful way. Starting from the gauge theory side, BMN
have been able to identify particular string states with gauge
invariant operators with large R-charge $J$, relating the energy of
the string states to the dimension of the operators. It has been shown
that these identifications are consistent with all planar
contributions \cite{santam} and non-planar diagrams for large $J$
\cite{sem, others}.

Although this is a very important development, it describes however only
one particular corner of the full range of gauge invariant operators
in SYM corresponding to large $J$. It is desirable therefore to extend
these results to wider class of operators. Some ideas about the
extension of AdS/CFT correspondence beyond the supergravity
approximation were actually suggested by Polyakov \cite{pol1} but
until recently they wasn't quite explored.

In an important recent developement, Gubser, Klebanov and Polyakov
\cite{gkp} (GKP) suggested another way of going beyond the supergravity
approximation. The main idea is to consider a particular configuration
of closed string in $\axs$ background executing a highly symmetric
motion. The theory of a generic string in this background is highly
non-linear, but  semiclassical treatment will ensure the
existence of globally conserved quantum numbers. AdS/CFT correspondence
will allow us to relate these quantum numbers to the dimensions of
particular gauge invariant operators in SYM. Gubser, Klebanov and
Polyakov investigated three examples:

a) rotating ``string soliton'' on $S^5$ stretched along the radial
direction $\r$ of AdS part. This case represents string states
carrying large R-charge and GKP have been able to reproduce the
results obtained from string theory in pp-wave backgrounds \cite{bmn}.

b) rotating ``string soliton'' on $AdS_5$ stretched along the radial
direction $\r$. This motion represents string states carrying spin
$S$. The corresponding gauge invariant operators suggested in
\cite{gkp} are important for deep inelastic scattering amplitudes as
discussed in \cite{kehag}.

c) strings spinning on $S^5$ and stretched along an angular direction
at $\r=0$.

All these states represent highly excited string states and since the
analysis is semiclassical, the spin $S$ is assumed quantized,

Shortly after \cite{gkp}, interesting generalizations appeared. In
\cite{fts, ts02, ts} a more general solution which includes the above
cases was investigated. Beside the solution interpolating between the
cases considered by GKP, the authors of \cite{fts, ts02, ts} have been
able to find a general formula relating the energy $E$, spin $S$ and
the R-charge $J$. In \cite{russo} a more general solution
interpolating between all the above cases was suggested. Further
progress has been made for the case of black hole $AdS$ geometry
\cite{petkou1} and confining AdS/CFT backgrounds \cite{petkou2,
alishah1}. The study of the operators with large R-charge and twist two
have been initiated in \cite{zar, kru}. Further interesting
developments of this approach can be found also in \cite{wadia,
alishah2}. We would note also the solution for circular configurations
suggested in \cite{minah}. The string soliton in this case can be
interpreted as a pulsating string.

From the preceeding discussion it is clear that the extension of the
semiclassical analysis to wider class of moving strings in $\axs$
background is highly desirable. The purpose of this note is to
consider a more general sigma model of closed strings moving in the above
background. One of the possibilities is to include closed string
with antisymmetric B-field turned on. We will consider the case of
short strings in B-field and will analyze the relation between the
conserved quantum numbers in the theory - in our case the energy $E$
and the spin $S$. 

The paper is organized as follows. In the next section we present a
brief review of rotating ``string soliton'' in $\axs$ background. In
Section 3, we start by considering short strings with B-field turned
on. In analogy with \cite{gkp}, we find the relation between the
energy $E$ and spin $S$. In Section 4, we study the quadratic
fluctuations around the classical solution of Sec. 3. We conclude with
summary and comments on the results and dicsuss some further
directions of development.

\section{Rotating strings in $\axs$}

In this section we will review the semiclassical analysis of rotating
strings \cite{gkp} following mainly \cite{fts, ts02, ts}.

The idea is to consider rotating and boosted closed ``string soliton''
stretched in particular directions of $\axs$ background. The general
form of the non-linear sigma model Lagrangian (bosonic part) can be
written as:
\eq{\label{1.1}
\L_B=\frac 12\sqrt{-g}g^{\a\b}\left[G^{(AdS_5)}_{ab}\p_\a X^a\, \p_\b
X^b + G^{(S^5)}_{nm}\p_\a X^n\,\p_\b X^m\right] 
}
The supergravity solutions of $\axs$ in global coordinates has the
form:
\eq{\label{1.2}
ds^2=-cosh^2\r\,dt^2+d\r^2+sinh^2\r\,d\Omega^2_{S^3} +d\psi^2_1+
cos^2\psi_1\left(d\psi^2_2+ cos^2\psi_2\,d\Omega^2_{\tilde S^3}\right)  
}
where,
\al{ \label{1.3} 
& d\Omega^2_{S^3}=d\b^2_1+ cos^2\b_1\left(d\b^2_2+ cos^2\b_2
\,d\b^2_3\right) \\
      \label{1.4}
& d\Omega^2_{\tilde S^3}=d\psi^2_3+ cos^2\psi_3\left(d\psi^2_4+ cos^2\psi_4
\,d\psi^2_5\right)       
}
(we have incorporated $R^2$ factor in the overall constant multiplying the
action, i.e. $1/\a\to R^2/\a$). Looking for solutions with conserved
energy and angular momentum, we assume that the motion is executed
along $\phi=\b_3$ direction on $AdS_5$ and $\theta=\psi_5$ direction
on $S^5$. Solution for closed string folded onto itself can be found
by making the following ansatz:
\al{ \label{1.5}
& t=\k\t;    \qquad \phi=\o\t;  \qquad \qquad \theta=\n\t \\
& \r=\r(\s); \qquad \b_1=\b_2=0; \qquad \psi_i=0 \quad (i=1,\dots 4) \notag
}
where $\o,\k$ and $\n$ are constants.
We note that one can always use the reparametrization invariance of
the worldsheet to set the time coordinate of space-time $t$ proportianal
to the worldsheet proper time $\t$. This means that our rod-like
string is rotating with constant angular velocity in the corresponding
spherical parts of $AdS_5$ and $S^5$.

Using the ansatz (\ref{1.5}) and the lagrangian (\ref{1.1}) one can
easily find the classical string equations, which in this case
actually reduces to the following equation for $\r(\s)$:
\eq{ \label{1.6}
\frac{d^2\r}{d\s^2}=\lb(\k^2-\o^2\rb)cosh\r\,\,sinh\r
}
and the Virasoro constraints take the form:
\eq{ \label{1.7}
{\r'}^2=\k^2 cosh^2\r -\o^2 sinh^2\r -\n^2.
}
One can obtain also the induced metric which turns out to be 
conformally flat as expected:
\eq{ \label{1.7i}
g_{\a\b}={\r'}^2\e_{\a\b},
}
where prime on $\r$ denotes derivative with respect to $\s$.
It is traightforward to obtain the constants of motion of the theory
under consideration\footnote{We follow the notatins of \cite{gkp} and
\cite{fts}}: 
\al{ \label{1.8}
& E=\sqrt{\l}\int\limits_0^{2\pi}\frac{d\s}{2\pi}\lb(-\frac{\p\L_B}
{\p\dot t}\rb)=\sqrt{\l}\k\int\limits_0^{2\pi}\frac{d\s}{2\pi}
cosh^2\r \\
     \label{1.9}
& S=\sqrt{\l}\int\limits_0^{2\pi}\frac{d\s}{2\pi}\lb(\frac{\p\L_B}
{\p\dot\phi}\rb)=\sqrt{\l}\o\int\limits_0^{2\pi}\frac{d\s}{2\pi}
sinh^2\r \\
     \label{1.10}
&  J= \sqrt{\l}\int\limits_0^{2\pi}\frac{d\s}{2\pi}\lb(\frac{\p\L_B}
{\p\dot\theta}\rb) =\sqrt{\l}\n
}
where we denoted $R^2/\a'$ as $\sqrt{\l}$. An immediate but important
relation following from (\ref{1.8}) and (\ref{1.9}) connect $E$ and
$S$:
\eq{ \label{1.11}
E=\sqrt{\l}\k+\frac\k\o S
}
Assuming that $\r(\s)$ has a turning point at $\pi/2$ and using the
periodicity condition\footnote{Since the closed string is folded onto
itself $\r(\s)=\r(\s+\pi)$.}, one can relate the three free parameters
$\k, \o$ and $\n$ as follows:
\eq{ \label{1.12}
coth^2\r_0=\frac{\o^2-\n^2}{\k^2-\n^2}=1+\e
}
where $\r_0$ is the maximal value of $\r(\s)$. A new parameter $\e$
introduced in (\ref{1.12}) is useful for studying the limiting cases
of short strings ($\e$ large) and long strings ($\e$ small). Using the
constraint (\ref{1.7}), the following expressions for the periodicity
condition and the constants of motion can be found:
\al{ \label{1.13}
& \sqrt{\k^2-\n^2}=\frac 1{\sqrt{\e}}
\,\, F\lb(\frac 12,\frac 12,1;-\frac 1\e\rb), \\
     \label{1.14}   
& E=\frac{\sqrt{\l}\k}{\sqrt{\k^2-\n^2}}\,\frac 1{\sqrt{\e}}
\,\, F\lb(-\frac 12,\frac 12,1;-\frac 1\e\rb), \\
      \label{1.15}
& S=\frac{\sqrt{\l}\o}{\sqrt{\k^2-\n^2}}\,
\frac 1{2\e^{3/2}}
\,\, F\lb(\frac 12,\frac 12,2;-\frac 1\e\rb), \\
}
where $F(\a,\b;\g;z)$ is the hypsegeometric function.

In the case of short strings $1/\e$ is very small. Straightforward
approximations lead to the relation between the energy 
$\E=E/\sqrt{\l}$ and the spin $\S=S/\sqrt{\l}$:
\eq{ \label{1.16}
\E\approx \sqrt{\n^2+\frac{2\S}{\sqrt{1+\n^2}}} +
\frac{\sqrt{\n^2+\frac{2\S}{\sqrt{1+\n^2}}}}
{\sqrt{1+\n^2+\frac{2\S}{\sqrt{1+\n^2}}}}\S
}
One can study several limiting cases. The simplest one is when $\n<<1$
and therefore $\S<<1$. In this case
\eq{ \notag
E^2\approx J^2+2\sqrt{\l}S
}
which can be interpreted as a string moving in flat space along a
circle with angular momentum $J$ and rotating in a plane with spin
$S$.

The second case is when the boost energy is much smaller than the
rotational energy: $\n^2<<\S$. The expression (\ref{1.16}) then
reduces to the flat space Regge trajectory \cite{gkp, fts}:
\eq{ \label{1.17}
\E\approx \sqrt{2\S} +\frac{\n}{2\sqrt{2\S}}
}
The last and most interesting case is when the boost energy is large,
i.e. $\n>>1$. In this case one can find:
\eq{ \label{1.18}
E\approx J + S +\frac{\l S}{2 J^2} +\dots
} 
which coincides with the leading order terms of BMN formula:
\eq{ \label{1.19}
E=J+\sum\limits_{n=-\infty}^\infty\sqrt{1+\frac{\l n^2}{J^2}}N_n 
+\O\lb(\frac 1{\sqrt{\l}}\rb)
}

Similar analysis can be done in the case of long strings. In this case
very interesting relations between the constants of motion were
found. We give below the results of \cite{gkp, fts, ts}

a) when $\n<< ln\,1/\e$
$$
E\approx S+\frac{\sqrt{\l}}\pi\,\,ln\,\lb(S/\sqrt{\l}\rb) +\frac{\pi J^2}
{2\sqrt{\l}\,\,ln\,\lb(S/\sqrt{\l}\rb)}
$$

b) when $\n>> ln\,1/\e$
$$
E\approx S+ J+\frac\l{2\pi^2 J}\,\,ln^2\lb( S/J\rb)
$$

We conclude this section by noting that  more general solutions can be
obtained by taking $\psi_1=\psi_1(\s)$ \cite{russo}. In this case the 
solutions interpolate smoothly between the cases described above.


\section{Classical solutions with B-field}

In this Section we consider rotating short strings in
the presence of B-field. The general sigma model action in this case
is given by (see for instance \cite{gsw}):
\al{  \label{2.1}
S &=-\frac{1}{4\pi\alpha'}\int\,d^2\sigma
\left\{\left[\sqrt{g}g^{\alpha\beta}G_{\mu\nu}(X)
\partial_\alpha X^\mu\partial_\beta X^\nu\right] \right. \notag \\
&+\varepsilon^{\alpha\beta}B_{\mu\nu}(X)\partial_\alpha X^\mu
\partial_\beta X^\nu+\alpha'\sqrt{g}\Phi(X)R^{(2)}
\left.\right\}
}
where the target space metric $G_{\m\n}$ is given by
(\ref{1.2},\ref{1.3},\ref{1.4}),  $g_{\a\b}$ is the worldsheet metric
and $\ep^{\a\b}$ is the 2-d antisymmetric tensor density.
 It is a simple exersice to check that the following ansatz is compatible
with the classical equations of motion:
\al{ 
& \b_3=\f=\o\t; \quad \b_1=\b_2=0; \qquad  t=\k\t   \notag \\  
& \psi_5=\th, \quad \th=\th(\s)=\th(\s+\pi); \qquad
 \r=\r(\s)=\r(\s+\pi) \label{2.2} \\
& B=b(\th)d\r\wedge d\f & \label{2.3} 
}
We will restrict our considerations to the simplest non-trivial case of
\eq{ \notag
H_{\th\r\f}= b =const
}
which means that $B_{\r\phi}$ is linear in $\theta$
\eq{ \label{2.4}
B_{\r\f}=\th(\s)\,b
}

The configuration defined by (\ref{2.2}) and (\ref{2.3}) describes
a "string soliton" stretched along $\r$ and
$\theta$, and moving on a circle along $\phi$ direction. With the choice
(\ref{2.2}-\ref{2.4}) the lagrangian reduces to
\al{ 
& \L=-\k^2 cosh^2\r+\lb(\p_\s\r\rb)^2+\o^2 sinh^2\r +\lb(\p_\s\th\rb)^2
+2\o B_{\r\f}(\th)\r' \notag \\
& =-\k^2 cosh^2\r +{\r'}^2+\o^2 sinh^2\r +{\th'}^2 +2\o b\r'\th. \label{2.5}
}

Straightforward calculaitons leads to the following string equation
of motion,
\al{ 
& \frac{d^2\r}{d\s^2}+b\o\frac{d\th}{d\s}
=(\k^2-\o^2)\, sinh\r\, cosh \r \label{2.6} \\
& \frac{d^2\th}{d\s^2} -b\o\frac{d\r}{d\s} =0. \label{2.7}
}
Note that in contrast to the previous cases and the more general solution
of \cite{russo}, the equations for $\r$ and $\theta$ are coupled, as
expected. Beside the equations of motion, we must ensure the conformal
invariance of the model, or in other words,  impose the Virasoro
constraints. The only non-trivial constraint can be read off from the 
lagrangian (\ref{2.5})
\eq{ \label{2.8}
\lb(\r'\rb)^2+\lb(\th'\rb)^2=\k^2cosh^2\r -\o^2 sinh^2\r
}
A straightforward check shows that the equations of motion (\ref{2.6},
\ref{2.7}) are compatible with (\ref{2.8}). 
Equations (ref{2.6}-\ref{2.8}) can also be derived from Nambu-Goto
action:
\eq{ \label{v1}
S_{NG}=\frac 1{4\pi\a'}\int d^2\s
\lb[\sqrt{\tilde g} +\ep^{\a\b} B_{\m\n}\p_\a X^\m\p_\b X^\n\rb]
}
where
\eq{ \label{v2}
\tilde g_{\a\b}=G_{\m\n}\p_\a X^\m\p_\b X^\n
}
is the induced metric on the worldsheet. Substituting the expressions
for $G_{\m\n}$ from (\ref{2.2}-\ref{2.4}) and using the ansatz
(\ref{2.2},\ref{2.3}) for $X^\m$ we find that
\al{
&\tilde g_{\t\t}=-\k^2 cosh^2\r+\o^2 sinh^2\r
\label{v2a} \\
&\tilde g_{\s\s}=\lb({\r'\,}^2+{\th'\,}^2\rb)
\notag
}
Substituting in the NG action (\ref{v1}), we get for the Lagrangian
\eq{ 
\label{v4}
\L\propto \sqrt{\k^2 cosh^2\r-\o^2
sinh^2\r}\sqrt{{\r'\,}^2+{\th'\,}^2}
- \o\b\th(\s)\r'(\s)
}
Equations of motion following from the Lagrangian (\ref{v4}) are
\eq{ \label{v5}
\frac d{d\s}\lb(\th'\frac{\sqrt{\k^2 cosh^2\r-\o^2 sinh^2\r}}
{\sqrt{{\r'\,}^2+{\th'\,}^2}}\rb) -b\o\r'=0
} 
and
\ml{ \label{v6}
\frac d{d\s}\lb(\r'\frac{\sqrt{\k^2 cosh^2\r-\o^2 sinh^2\r}}
{\sqrt{{\r'\,}^2+{\th'\,}^2}}\rb) 
-\frac{(\k^2-\o^2)sinh\r cosh\r\sqrt{{\r'\,}^2+{\th'\,}^2}}
{\sqrt{\k^2 cosh^2\r-\o^2 sinh^2\r}}\\
+b\o\th'=0
}
Now the induced metric on the worldsheet can always be put in the
conformal gauge. This requires $g_{\t\t}=-g_{\s\s}$. 
Thus we require that
\eq{ \label{v7}
{\r'\,}^2+{\th'\,}^2=\k^2 cosh^2\r-\o^2 sinh^2\r
}
It can be readily seen that (\ref{v5}) and (\ref{v6}) reduce to
(\ref{2.7}) and (\ref{2.6}) respectively.

Before we proceed further, we
have to first eliminate the $\theta$ dependence in (\ref{2.8}) so that
to be able to express all the quantities in terms of $\r$ only
(actually we need $\r'$ only).
Integrating (\ref{2.7}) once  we find:
\eq{ \label{2.9}
\th'=\a+b\o\r(\s)
}
As in the previous Section, we make two assumptions. First, we consider
$\r$ periodic in $\s$ with turning point at $\pi/2$. Secondly, we choose
the minimal value of $\r$ at $\s=0$ to be vanishing. With this choice
the integration constant $\a$ is fixed to be 0:
\eq{ \label{2.10}
\th'=b\o\r(\s)
}
Now we can substitute for $\theta'$ into the Virasoro constraint
(\ref{2.8}) to find:
\eq{ \label{2.11}
{\r'}^2 =\k^2 cosh^2\r -\o^2 sinh^2\r -b^2\o^2\r^2
}
Since we consider rotating short strings, there are restrictions on the
range of the parameters  $\k$ and $\o$, coming from
the periodicity of $\r$. At the turning point ($\s=\pi/2$) $\r'$
vanishes. This gives the relation between $\k , \o$ and $b$:
\eq{ \label{2.12}
0=\k^2 cosh^2\r_0-\o^2 sinh^2\r_0-b^2\o^2\r^2_0
}
or
\eq{ \label{2.13}
coth^2\r_0=\frac{\o^2}{\k^2}\lb( 1+ b^2\frac{\r^2_0}
{sinh^2\r_0}\rb)
}
For short strings $\r_0$ is very small, so we can use the following
approximation:
\eq {
 sinh^2\,\r_0\approx \r^2_0 \label{2.14} 
}
and then (\ref{2.13}) becomes
\eq {
coth^2\r_0\approx \frac{\o^2}{\k^2} \lb(1+ b^2\rb)
\label{2.15}
}
In order for (\ref{2.15}) to be valid for generic $b$ the ratio
$\o^2/\k^2$ must be large (as $coth^2\r_0$ is large for small $\r_0$).
As in Section 2, we define a new parameter $\e$ through
\eq{ \label{2.18}
\frac{\o^2}{\k^2} = 1+\e
}
Within this approximation  eq. (\ref{2.15}) can be written as:
\al{ 
& cosh^2\r_0-\frac{\o^2}{\k^2} sinh^2\r_0 - b^2\frac{\o^2}{\k^2}\r^2_0
\notag \\
& \approx 1-\lb[b^2+\e(1+b^2)\rb]\r^2_0=0
}
so that:
\eq{ \label{2.19}
\r_0\approx \frac{1}{\sqrt{\x}}\,\,; \qquad \x=b^2+\e(1+b^2) .
}


We expect that the periodicity condition on $\r(\s)$ will give us
a relation between $\k$ and $\x$. Indeed, using the approximation
(\ref{2.14}) we find:
\eq{ \label{2.20}
d\s\approx\frac{d\r}{\k\sqrt{1-\x\r^2}}
}
or, integrating over $\s$
\eq{ \label{2.21}
2\pi =\frac 4\k\int\limits_0^{\r_0}\frac{d\r}{\sqrt{1-\x\r^2}}
}
In other words, the periodicity condition reduces to:
\eq{ \label{2.22}
1\approx sin\frac{\pi}{2}\k\sqrt{\x}
}
and therefore
\eq{ \label{2.23}
\k\approx\frac{1}{\sqrt{\x}}=\frac{1}
{\sqrt{b^2+\e(1+b^2)}}
}
Now we proceed with the constants of motion. The boost energy can be
defined in a standard way:
\al{ \label{2.24}
S&=\sqrt\l\int\limits_0^{2\pi}\frac{d\s}{2\pi}
\lb(\frac{\p\L}{\p\dot\f}\rb)=\sqrt\l\o\int\limits_0^{2\pi}\frac{d\s}{2\pi}
sinh^2\r .
}
It is useful to define a $\l$ independent variable $\S=S/\sqrt\l$.
In short string approximation the expression for $\S$ greately
simplifies and takes the form:
\eq{ \label{2.25}
\S\approx\frac{2\o}{\pi\k}\int\limits_0^{\r_0}\frac
{\r^2\,d\r}{\sqrt{1-\x\r^2}}.
}
The integral in (\ref{2.25}) can be easily evaluated and the final
expression for $\S$ in terms of the constant parameters of the model
is
\eq{ \label{2.26}
\S\approx \frac\o{2\k\x\sqrt\x}.
}
We see that the expression for $\S$ is similar to that obtained in
\cite{gkp, fts, ts}, but with "deformed" parameter $\e\to\x$. One
can use now the relation between $\k$ and $\o$ (\ref{2.18})
(and taking into account that $\k\approx 1/\sqrt{\x}$) to obtain
\eq{ \label{2.27}
\o^2\approx (1+\e)\frac 1\x.
}
For short strings, $\e$ is very large and therefore
\eq{ \label{2.28}
\S\approx\frac 1{2\e(1+b^2)^{3/2}}
}
or
\eq{ \label{2.29}
\e\approx\frac 1{2\S(1+b^2)^{3/2}}>>1 .
}
Note that in this case $\S$ has to be very small: $\S<<1$.

Since in AdS/CFT correspondence, the dimensions of the operator in the
gauge theory side are determined by the energy, we want to find
a relation between the spin $\S$ and the energy. The later can be
obtained from the lagrangian (\ref{2.5}) in a stantard way
\eq{ \label{2.30}
E=\sqrt\l\int\limits_0^{2\pi}\frac{d\s}{2\pi}
\lb(-\frac{\p\L}{\p\dot t}\rb)
=\sqrt\l\k\int\limits_0^{2\pi}\frac{d\s}{2\pi}cosh^2\r
}
Defining $\l$ independent quantity
\eq{ \label{2.31}
\E=\k\int\limits_0^{2\pi}\frac{d\s}{2\pi}cosh^2\r
}
and using (\ref{2.24}) we obtain the same relation between $\E$ and
$\S$ as in \cite{gkp, fts}
\eq{ \label{2.32}
\E=\k+\frac\k\o\S
}
It is not necessary to calculate $\E$ explicitly but instead simply
use the above relation between $\E$ and $\S$. The final expression
then takes the form:
\eq{ \label{2.33}
\E=\frac{\sqrt{2\S\sqrt{1+b^2}}}{\sqrt{1+2b^2\sqrt{1+b^2}\S}}
+ \frac{\sqrt{2\S(1+b^2)^{3/2}}}
{\sqrt{1+2\S(1+b^2)^{3/2}}}
}
Since, when $b=0$ (\ref{2.33}) reproduces the corresponding case
of \cite{gkp, fts}
\eq{ \label{2.36}
\E=\sqrt{2\S}+\frac{\sqrt{2\S}}{\sqrt{1+2\S}},
}
one can interpret (\ref{2.33}) as a smooth deformation of the latter. For
small boost energy we recover again the flat space Regge trajectory
\eq{ \label{2.35}
\E\approx \sqrt{2\S}
}

%
%

In the case of small $\r_0$ the model looks as a point-like string
moving in the presence of closed string B-field. In this approximation
the solutions for $\r$ and $\theta$ are:
\al{ 
& \r\approx\frac 1{\sqrt{b^2+\e(1+b^2)}} sin\s \label{2.37} \\
&\th\approx \th_0-\frac 1{\sqrt{b^2+\e(1+b^2)}} cos\s
}
and since $\e$ is very large the amplitudes of $\r$ and $\theta$ are
very small. Again, when $b=0$, we find the solution of \cite{gkp, fts, ts}
\eq{ \label{2.39}
\r\approx \frac 1{\sqrt\e} sin\s
}

We conclude this Section with the following remark. One can choose the
B field to be oriented along ($\th,\f$) plane and depending on
$\r$. If we impose the condition of constancy of the field strength,
we end up with
\eq{ \label{2.40}
B_{\th\f}=b\r(\s).
}
The eqs. of motion in this case are slightly modified
\al{
& \r''-b\o\th'=\lb(\k^2-\o^2\rb) sinh\r\, cosh\r \label{2.41} \\
& \th''+b\o\r'=0 \label{2.42}
}
with the same constraint
\eq{ \label{2.43}
\lb(\r'\rb)^2+\lb(\th'\rb)^2=\k^2 cosh^2\r-\o^2 sinh^2\r .
}
Eliminating $\th$ from (\ref{2.43}) we arrive at the same expression
for $\r'$ as in (\ref{2.11})
\eq{ \label{2..44}
\lb(\r'\rb)^2+b^2\o^2\r^2    =\k^2 cosh^2\r -\o^2 sinh^2\r 
}
From here on the analysis proceeds in the very same way as in the
above and with the same conclusions.

\section{Quadratic fluctuations}

In this Section we would like to study the quadratic fluctuations around
the classical solutions (\ref{2.3},\ref{2.4}). The study of the
fluctuations around given particular string configuration means
semiclassical approximation of the theory and therefore it allows us
to study the leading quantum corrections to the energy spectrum.

Let us start with some string configuration $\bar X^\m$ satisfying the
classical equations of motion. A standard method for studying small
fluctuations around $\bar X^\m$ is the use of Riemann normal coordinates. Our
starting point will be the action:
\eq{ \label{3.1}
S=-\frac 1{4\pi\a'}\int d^2\s\lb\{
\sqrt{g}g^{\a\b}G_{\m\n}\p_\a X^\m \p_\b X^\n +
\ep^{\a\b} B_{\m\n}\p_\a X^\m \p_b X^\n \rb\}
}
The expansion of the entries of (\ref{3.1}) is given by (see for
instance \cite{gsw})  
\al{
& \p_\a X^\m=\p_\a\bar X^\m + D_\a\x^\m+\frac 13 R^\m_{\l\k\n}\x^\l
\x^\k\p_\a\bar X^\n+\cdots \notag \\
& G_{\m\n}(X)= G_{\m\n}(\bar X)-\frac 13 R_{\m\r\n\k}\x^\r\x^\k 
+ \cdots \notag \\
& B_{\m\n}(X)= B_{\m\n}(\bar X)+D_\r B_{\m\n}(\bar X)\x^\r+
\frac 12 D_\l D_\r B_{\m\n}(\bar X)\x^\l\x^\r - \notag \\
&\qquad \qquad - \frac 16 R^\l_{\r\m\k}(\bar X)B_{\l\n}(\bar X)\x^\r\x^\k
+\frac 16 R^\l_{\r\n\k}B_{\l\m}(\bar X)\x^\r\x^\k+\cdots
\notag
}
where
\eq{ 
D_\a\x^\m=\p_\a\x^\m+\G^\m_{\l\n}\x^\l\p_\a\bar X^\n  \notag
}
and $R^\m_{\n\r\l}$ is the usual Riemann tensor for the target space.
Now we can use the above expansion to obtain the quadratic part of 
(\ref{3.1}) 
\eq{ \label{3.2}
S^{(2)}= S^{(2)}_G + S^{(2)}_B \notag
}
where
\al{
& S^{(2)}_G=-\frac 1{2\pi\a'}\int d^2\s\,\sqrt{g}g^{\a\b}
\lb[ G_{\m\n}(\bar X) D_\a\x^\m D_\b\x^\n + A_{\m\n;\a\b}\x^\m\x^\n
\rb] \label{3.3} \\
& S^{(2)}_B = -\frac 1{2\pi\a'}\int d^2\s\, \ep^{\a\b} \lb[
\p_\a\bar X^\d H_{\d\m\n}\x^\n D_\b\x^\m +
\frac 12 D_\l H_{\d\m\n}\x^\l\x^\d\p_\a\bar X^\m\p_\b\bar X^\n
\rb] \label{3.4}
}
The explicit form of $ A_{\m\n;\a\b}$ is given by\footnote{The
expressions are analogous to those in \cite{dgts} for the open
superstring and the mass term  $A_{\m\n;\a\b}$ term was also obtained 
there.}:
\eq{ \label{3.5} 
A_{\m\n;\a\b}=R_{\m\d\n\k}(\bar X)\p_\a\bar X^\d \p_\b\bar X^\k
}
One can rewrite (\ref{3.3}) in Lorentz frame by using vielbeins
 defined by
\al{
& G_{\m\n}=E^A_\m E^B_\n \,\e_{AB} \notag \\
& \x^A=E^A_\m\x^\m .\notag 
}
After some strightforward  calculations the fluctuation part
$S^{(2)}_G$ is found to be
\eq{
S^{(2)}_G=-\frac 1{4\pi\a'}\int d^2\s \sqrt{g}
\lb[g^{\a\b}\e_{AB}\D_\a\x^A\D_\b\x^B+
\lb(g^{\a\b}R_{ACBD}Y^C_\a Y^D_\b\rb)\x^A\x^B\rb]
\label{3.7}
}
where
\al{ \label{3.7a}
&\D_\a\x^A=\p_a\x^A+\o^{AB}_\a\x^B\\
& Y^A_\a=E^A_\m\p_\a\bar X^\m \notag
}
and the projected spin connection is given by
\al{
& \o^A_{B\m}=E^A_\n\lb(\p_\m E^\n_B+\G^\n_{\m\l} E^\l_B\rb) \notag \\
& \o^{AB}_\a=\o^{AB}_\m\p_\a\bar X^\m . \notag 
}
It is straightforward to show that the only non-vanising projected
spin connection coefficients are\footnote{The spin connection
coefficients and the projected vielbeins were also used in
\cite{fts}.}
\eq{ \label{3.12}
\o^{01}_0=\k\, sinh\r\,\,, \qquad \o^{21}_0=\o\, cosh\r.
}
and it is also useful to introduce projected vielbeins
\eq{ \notag
e^A_\a=E^A_\m\p_\a\bar X^\m
}
\al{ 
& e^0_0=\k\, cosh\r\,; & e^1_1 =\r'\, \notag \\
& e^2_0=\o\, sinh\r\,; & e^3_1 =\th'\, \label{3.13}
}
In these notations the massive term for the non-trivial $AdS$ part 
becomes
\eq{ \label{3.14}
A_{AB}=g^{\a\b}\e_{CD}e^C_\a e^D_\b\,\e^{AB}-
g^{\a\b}e^A_\a e^B_\b
}
We will use the fact that the induced metric is conformally flat to
replace $g_{\a\b}$ by $\e_{\a\b}$. The first term in (\ref{3.14}) then
becomes
\al{
\e^{\a\b}\e_{CD}e^C_\a e^D_\b\,\e^{AB} & =
2\lb({\r'\,}^2+{\th'\,}^2\rb)\e^{AB} \notag \\
& = 2\lb({\r'\,}^2+b^2\o^2\r^2\rb)\e^{AB}. \label{3.15}
}
The total mass term then becomes
\al{
A_{AB}\x^A\x^B=&\lb(\r'\x^0-\th'\x^3\rb)^2+2\lb({\r'\,}^2+{\th'\,}^2\rb)
\lb(-{\x^0\,}^2+{\x^2\,}^2\rb) \notag \\
&+\lb(\k\, cosh\r\x^0-\o\, sinh\r\x^2\rb)^2+\lb(\r'\,\x^1-\th'\,\x^3\rb)^2.
\label{3.16}
}
Now we give the final expression for the quadratic fluctuations of
$S^{(2)}_G$:
\al{
S^{(2)}_G=& -\frac 1{4\pi\a'}\int d^2\s\lb[
\lb(\p_0\x^0+\k\,sinh\r\,\x^1\rb)^2
-\lb(\p_0\x^1+\k\,sinh\r\,\x^0-\o\,cosh\r\,\x^2\rb)^2 \right. \notag \\
& -\lb(\p_0\x^2+\o\,cosh\r\,\x^1\rb)^2 -\lb(\p_1\x^0\rb)^2 
 +\lb(\p_1\x^1\rb)^2 + \lb(\p_1\x^2\rb)^2 +\lb(\p_1\x^3\rb)^2
\notag \\
& +\lb(\r'\x^0-\th'\x^3\rb)^2 
 +2\lb({\r'\,}^2 +{\th'\,}^2\rb)
\lb(-{\x^0\,}^2+{\x^2\,}^2\rb) \notag \\
& \left. + \lb(\k\,cosh\r\,\x^0-\o\,sinh\r\,\x^2\rb)^2  
 + \lb(\r'\x^1-\th'\x^3\rb)^2\rb] \label{3.17}
}

Now we proceed with the quadratic fluctuations for B-part of the
action (\ref{3.4}). In Lorentz frame it takes the form
\ml{
S^{(2)}_B= -\frac 1{4\pi\a'}\int d^2\s\,\ep^{\a\b}
\lb[\lb(\p_\a\bar X^\d H_{\d\m\n}\,E^\n_A E^\m_B\rb)\x^A\D_\b\x^B
\right. \\
+\left. \frac 12\lb(D_\l H_{\d\m\n}\p_\a\bar X^\m\p_\b\bar
X^\n\rb)\x^A\x^B\rb] \notag
}
Lengthy but straightforward calculations lead to the following final
result for $S^{(2)}_B$
\ml{ \label{3.18}
S^{(2)}_B= -\frac 1{4\pi\a'}\int d^2\s\lb[\o\lb(\x^\r\p_1\x^\f
-\x^\f\p_1\x^\r\rb)+
\r'b\lb(\x^\f\p_0\x^\th-\x^\th\p_0\x^\f\rb) \right. \\
\left. +b\,\th'\lb(\x^\r\p_0\x^\f-\x^\f\p_0\x^\r
+\k\,sinh\r\,\x^t\x^\f\rb)\rb]
}

The quadratic fluctuations for G-part of the action, $S^{(2)}_G$,
looks very similar to those obtained in \cite{fts,ts}, but the
additional part $S^{(2)}_B$ adds new feature. Due to (\ref{3.18}) the
equations for different modes are coupled and therefore the
presence of B-field  will drastically change the energy spectrum. We
will address the detailed study of this issue to future investigations.

\section{Conclusions}

In present paper we considered a rotating ``string soliton'' in the
presence of NS-NS B-field. Using suitable choice of the string
configuration and constant field strength for $B_{\m\n}$, we analyse (in
the case of short strings) the classical values of the energy $\E$ and
the spin $\S$ of the system. The obtained values for the constants of
motion deviate smoothly from those obtained without B-field:
\eq{ \label{c1}
\E=\frac{\sqrt{2\S\sqrt{1+b^2}}}{\sqrt{1+2b^2\sqrt{1+b^2}\S}}
+ \frac{\sqrt{2\S(1+b^2)^{3/2}}}
{\sqrt{1+2\S(1+b^2)^{3/2}}}
}
\eq{ \label{c2}
\S\approx \frac\o{2\k\x\sqrt\x}
}

One can analyse two limiting cases for the value of the field strength
$b$.

a) First case obviously is when $b$ is very small (comapared to
$\S$). In this case the contribution of the B-field can be neglected
and the smooth limit $b\to 0$ reduces to the case studied in \cite{gkp, fts}.

b) The second possibility is when $b\S$ is very large (note that
according to (\ref{2.29}) $\S$ is small). From (\ref{c1})
then one can find that
\eq{ \label{c3}
\E=\frac{\sqrt{2\S\sqrt{1+b^2}}}{\sqrt{1+2b^2\sqrt{1+b^2}\S}}
+ \frac{\sqrt{2\S(1+b^2)^{3/2}}}
{\sqrt{1+2\S(1+b^2)^{3/2}}}\propto 1+\frac 1{b}+\dots
}
and the dependence on $\S$ drops out from the leading contributions.

In Section 4 we considered the fluctuations around the classical
solutions of Section 3. The contributions due to the presence of NS-NS
antisymmetric B-field couple the fluctuation modes in a non-trivial
way. This indicates changes in the quantum corrections to the
energy spectrum compared to the case without B-field.

This paper is incomplete in several ways. First of all we
analysed the case of short strings only.  Actually the long string
case investigated in \cite{gkp} leads to a more involved relation
between the energy and the spin:
\eq{
E=S+\frac {\sqrt{\l}}\pi \log(S/\sqrt{\l})+\dots
\notag
}
In our case the energy $\E$ and the spin $\S$ are defined by
\al{
& E=\frac{\k\sqrt{\l}}{2\pi}\int\limits_0^{\r_0}
\frac{cosh^2\r\,d\r}
{\sqrt{\k^2cosh^2\r-\o^2 sinh^2\r-b^2\o^2\r^2}}\notag \\
& S=\frac{\o\sqrt{\l}}{2\pi}\int\limits_0^{\r_0}
\frac{sinh^2\r\,d\r}
{\sqrt{\k^2cosh^2\r-\o^2 sinh^2\r-b^2\o^2\r^2}}\notag
}
Although we do not have explicit solution so far, one can speculate that
since the integral is dominated by the contributions around the
turning point $\s=\pi/2$, the behavior of $\E$ and $\S$ should be
qualitatively the same. One can expect then that the relation between
$\E$ and $\S$ will be approximately the same, however the impact of
B-field on this speculation remain unclear and needs more rigorous
study. We will return to this question in the near future. 

The second question we did not consider in this paper is the
fermionic part. This however is important for several reasons. Besides
supersymmetry, we should note that the fermions are important also
for divergence cancellations of the 2d (induced metric)
curvature\footnote{See for detailed study in the case of open
superstrings \cite{dgts}}. Indeed, the contribution of the
fluctuation modes to the logarithmic divergences is proportional to
the trace of the mass matrix $A_{AB}$ (\ref{3.5}) 
We expect that these divergences will be cancelled by the
contributions coming from the fermions in the same way as in
\cite{fts} and the theory then will remain conformally invariant. This
point however needs a rigorous clarification which we leave for
separate study.

The last, but maybe the most important question is the identification
of the energy with the dimension of certain operators from SYM theory
side. Since this is an important question we will return to this issue
in the near future.

One more remark is in order. After suitable redefinitions, the
equations for the fluctuations can be brought to a form very similar
to those obtained in \cite{myers}. One can wonder if there is a deeper
relation beyond the similarity between the case studied here and
pp-wave limit of Pilch-Warner geometries studied in \cite{myers}.

We hope that further studies of rotating strings in presence of
NS-NS B-field will shed more light on AdS/CFT correspondence.

\vspace*{.8cm}

{\bf Acknowledgements:} We would like to thank R.Parthasarathy for
fruitful discussions. R.R. would like to thank Simon Fraser
University for warm hospitality. This work has been partially supported by an
operating grant from the Natural Sciences and Engineering Research Council
of Canada.

\end{document}